\documentclass{emulateapj}

\newcommand{\simless}{\mathbin{\lower 3pt\hbox
      {$\rlap{\raise 5pt\hbox{$\char'074$}}\mathchar"7218$}}} 
\newcommand{\simgreat}{\mathbin{\lower 3pt\hbox
     {$\rlap{\raise 5pt\hbox{$\char'076$}}\mathchar"7218$}}} 

\slugcomment{\today}

\shorttitle{The mm-colors of a young binary disk system in the ONC}
\shortauthors{Ricci et al.}

\begin{document}


\title{The mm-colors of a young binary disk system in the Orion Nebula Cluster}


\author{L. Ricci, L. Testi}
\affil{European Southern Observatory, Karl-Schwarzschild-Strasse 2,
D-85748 Garching bei M$\ddot{u}$nchen}

\author{J. P. Williams}
\affil{Institute for Astronomy, University of Hawaii, 2680 Woodlawn Drive, 96822 Honolulu, HI, USA}

\author{R. K. Mann}
\affil{National Research Council Canada, Herzberg Institute of Astrophysics, 5071 West Saanich Road, Victoria, BC V9E 2E7, Canada}

\and

\author{T. Birnstiel}
\affil{Max-Planck-Institut f\"ur Astronomie, K\"onigstuhl 17, 69117 Heidelberg, Germany}


\email{lricci@eso.org}


\begin{abstract}

\noindent We present new EVLA continuum observations at 7~mm of the 253-1536 binary disk system in the Orion Nebula Cluster. The measured fluxes were combined with data in the sub-mm to derive the millimeter spectral index of each individual disk component.
We show how these observations can be used to test the models of dust evolution and early growth of solids in protoplanetary disks. Our analysis indicates that the disk with lower density and higher temperature hosts larger grains than the companion disk. This result is the opposite of what predicted by the dust evolution models. The models and observational results can be reconciled if the viscosity $\alpha$-parameter differs by more than a factor of ten in the two disks, or if the distribution of solids in the disks is strongly affected by radial motions. 
This analysis can be applied to future high-angular resolution observations of young disks with EVLA and ALMA to provide even stronger observational constraints to the models of dust evolution in protoplanetary disks. 
 
\end{abstract}


\section{Introduction}
\label{sec:intro}

\noindent Young circumstellar disks around pre-Main Sequence (PMS) stars are thought to host the birth of planets.
According to the core accretion scenario, the formation of planets involves a variety of physical mechanisms from the coagulation of sub-$\mu$m sized particles up to the gas-accretion phase leading to the build up of gas giants. 
One of the most critical steps which is left to be explained is how this growth of solids proceeds from mm/cm-sized dust grains to km-sized rocks, called planetesimals. This is crucial because it is from the gravitational-driven collision of these planetesimals that both the rocky Earth-like planets and the rocky cores of giant planets are ultimately supposed to be formed. Different mechanisms have been proposed in the literature to explain the formation of planetesimals. In general, these mechanisms induce the local accumulation of particles with sizes of $\sim 1-100$~mm at densities which are high enough to make these clumps unstable to gravitational collapse \citep[][]{Chiang:2010}.
It is therefore very important to well characterize the early phases of coagulation of small dust grains in order to set the right initial conditions for the mechanisms governing the formation of planetesimals.

Recently, sophisticated models of dust evolution in disks including several effects like dust coagulation, fragmentation and radial motions, have been built \citep{Brauer:2008,Birnstiel:2010a}. 
Continuum observations in the (sub-)millimeter constrain fundamental parameters of the dust population in the disk, e.g. the total dust mass, the radial-dependent dust surface density, and the size distribution of dust grains at sizes of about $\sim 0.1-10$~mm \citep[see][]{Williams:2011}. These observational constraints can be used to test the models of dust evolution and shed light onto the first stages of planetesimal formation \citep{Birnstiel:2010b}.  

The recent initial upgrades of the Very Large Array into the Expanded Very Large Array \citep[EVLA;][]{Perley:2011} allowed a significant increase in continuum sensitivity at long wavelengths.  
We present new EVLA observations at about 7~mm of the 253-1536 binary system in the Orion Nebula Cluster. The projected angular separation of the two PMS stars is about 1.1~$arcsec$ which corresponds to a projected physical separation of 460~AU at the ONC distance of $\sim$ 420~pc~\citep{Kraus:2009}. Optical images with the Hubble Space Telescope (HST) has revealed the presence of a large disk seen in absorbtion around the east companion, 253-1536a. \citet{Mann:2009} observed the binary at 0.88~mm with the Sub-Millimeter Array (SMA) and estimated for the 253-1536a disk a mass of about 0.07~$M_{\odot}$ which makes this disk the most massive ever observed in the ONC. They also detected a fainter disk around the other PMS star, 253-1536b, and found a ratio for the two disk masses of about 4.  

\section{Observations}
\label{sec:obs}

\subsection{EVLA}
\label{sec:obs_EVLA}

\noindent We obtained new EVLA data under the project 10B-102, with a total observing time of 6 hours.
At the time of observations the EVLA was in the C configuration, and provided a total bandwidth of 256~MHz, that we centered at the frequency of 43.280~GHz (6.9~mm) in the Q-band. 

Passband and absolute flux calibration was performed through observations of the QSO 3C147 (J0542$+$4951).
Amplitude and phase calibration was obtained by observing the QSO J0607$-$0834. The measured 6.9~mm-flux density of J0607$-$0834 range from about 1.9~Jy to 2.1~Jy during the three observing runs, thus showing a discrepancy in the fluxes which is within the $\sim 10\%$-uncertainty on the absolute flux, as typically estimated for EVLA observations in the Q-band. 

The raw visibilities were calibrated using the Common Astronomy Software Applications (\textit{CASA}) package. Maps of the continuum emission were derived by adopting natural weighting to maximize the observations sensitivity, and photometry was obtained through Gaussian fitting in the image plane using the \texttt{imfit} task in \textit{CASA}. The resulting synthesized beam has sizes of about $0.77\arcsec \times 0.55\arcsec$. The measured rms-noise of the final map is about 35~$\mu$Jy.   

\subsection{X-Shooter}
\label{sec:X-Shooter}

\noindent We observed 253-1536 with XShooter at the ESO-VLT on  
September 27,
2010. We obtained complete spectra from 330 through 2500~nm at a  
spectral resolution
in the range 5000-9000. We aligned the $\sim 11\times 1.0^{\prime 
\prime}$~arcsec slit
along the binary position angle to observe the two components  
simultaneously.
We performed an on-slit dithering pattern with 8 exposures of 190s  
each, to avoid
saturation on the brightest component and to efficiently remove sky  
emission, bad pixels
and cosmic ray hits. We reduced the data using the standard XShooter  
pipeline
recipes, and IRAF for extraction of the 1D spectra  
and subsequent analysis.
We used spectra of LTT~1020 and Hip~025028 (B3V) to correct for instrumental response and telluric absorption. The  
spectrum of the bright
component is detected with very high signal-to-noise across the whole  
spectral range, while the
faint component of the binary system is only detected at K and H bands  
and cannot be easily
separated from the glow of the other component at shorter wavelengths.

We detect numerous absorption and emission features in the two  
spectra. Nevertheless, for
the purpose of this Letter we just analysed the spectra to attempt a  
spectral classification
of the two components of the binary. We compared our  
spectra with those from the
spectral libraries of \citet{Allen:1995} and \citet{Wallace:1997}. We find that the
bright stellar component of the binary, i.e. 253-1536b, is consistent with previous  
classification as an M2 star both based on the optical and near  
infrared spectrum. The faint component, 253-1536a, does not show  
prominent
absorption lines. In particular, it is laking CO, Fe, Al and Ca  
features that are commonly observed in
K-band spectra of late type dwarfs. Overall the spectrum seems to be  
consistent with heavily veiled and extincted late F and G types. In this paper we will consider a spectral type of G2 for the 253-1536a PMS star.

\section{Results}
\label{sec:res}

\begin{figure}
 \includegraphics[scale=1.0]{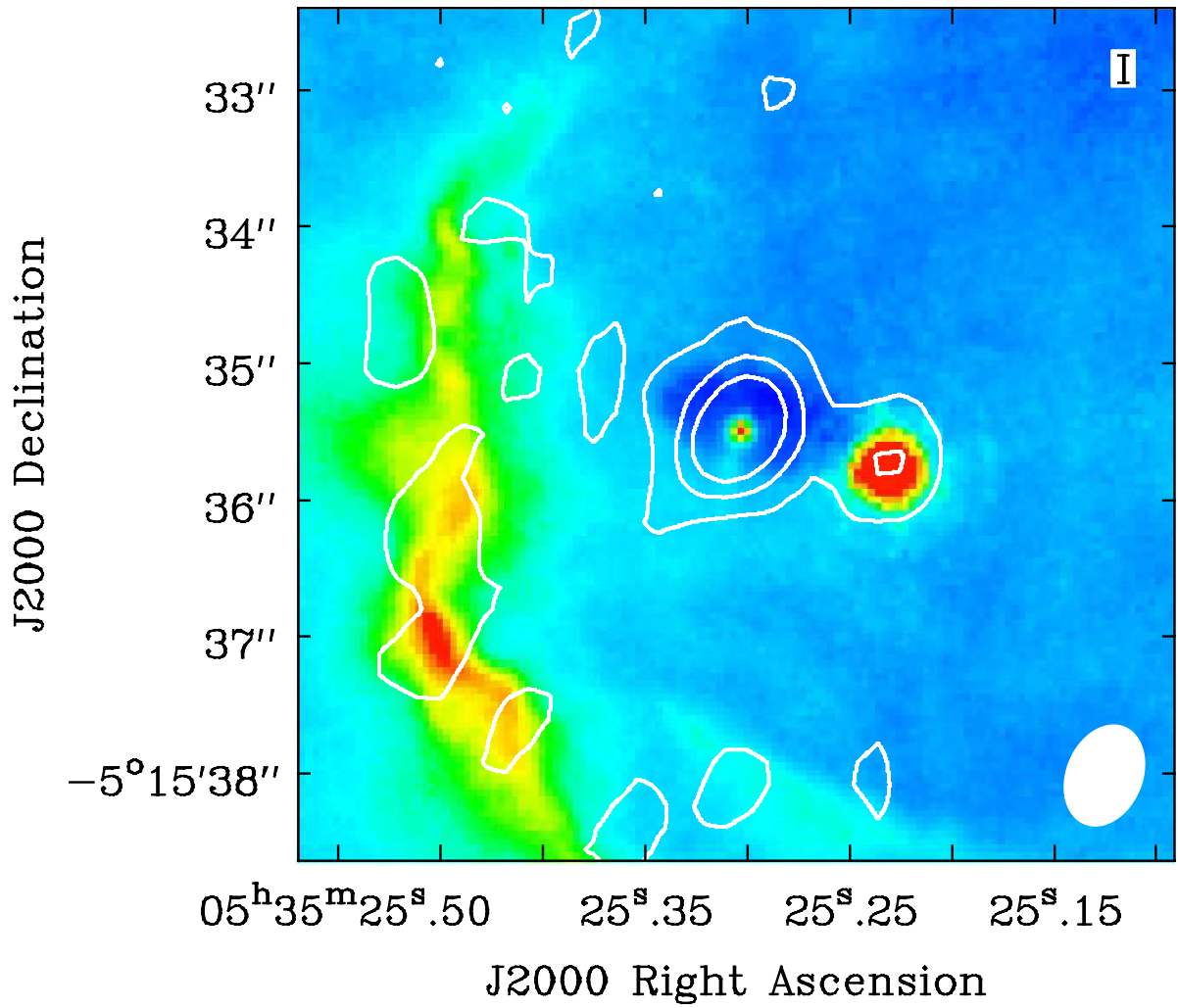}
\caption{Map of the 253-1536 binary system. In color the HST image in the H$\alpha$ F658N filter \citep{Ricci:2008}, whereas white contours represent the EVLA continuum data at 6.9~mm. Contour lines are drawn at 3, 6 and 9$\sigma$, where $\sigma = 0.035$~mJy/beam. The white ellipse in the lower right corner indicates the size of the EVLA synthesized beam.}
\label{fig:map2}
\end{figure}

\noindent The contours in Figure~\ref{fig:map2} show the EVLA map at 6.9~mm superimposed to the HST image in H$\alpha$.
Both the disks are clearly detected at 6.9~mm, and the measured fluxes are listed in Table~\ref{tab:obs}, together with the ones measured with the SMA array at 0.88~mm \citep{Mann:2009}.
Continuum observations of young circumstellar disks in the sub-mm probe the emission from dust particles. At longer millimeter wavelengths the dust emission can be contaminated by ionized gas, with ionization driven by X-ray/UV-radiation generated either by nearby young massive stars, e.g. in the Trapezium Cluster \citep[][]{Williams:2005}, or by the PMS star at the center of the disk, as seen for disks in low mass SFRs \citep[][]{Rodmann:2006}.    

\begin{figure}
 \includegraphics[scale=0.7]{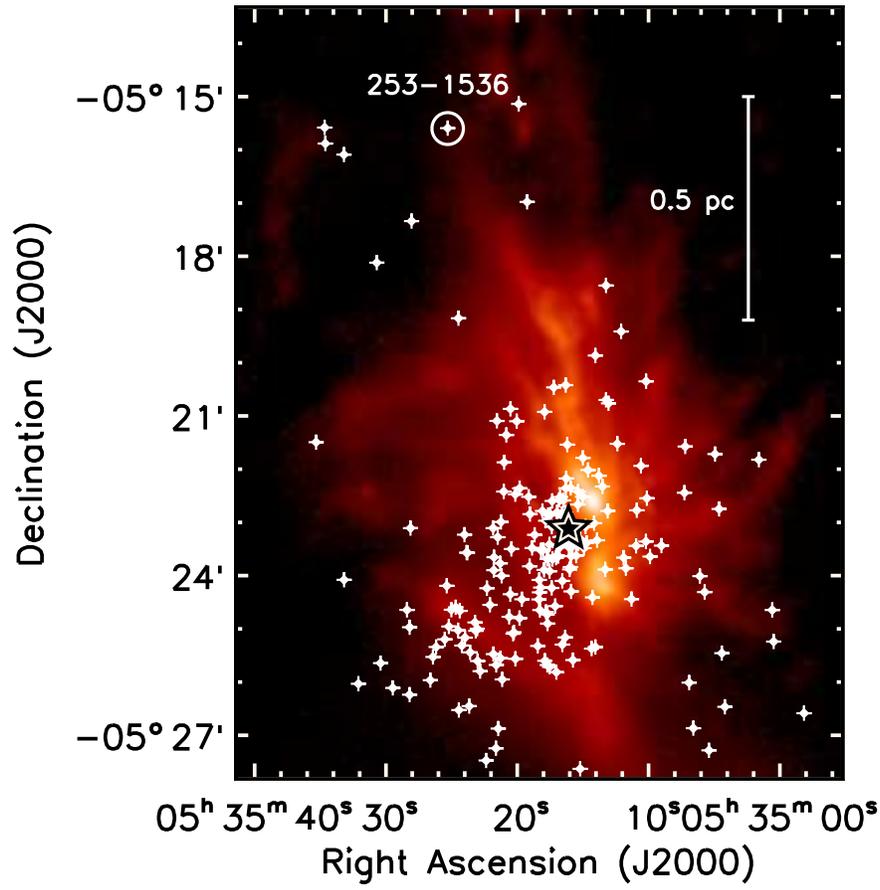}
\caption{ONC map at 450~$\mu$m with SCUBA/JCMT \citep{Johnstone:1999}. White crosses indicate the HST-identified proplyds, whereas the star marks the location of $\theta^1$ Ori C. The location of the 253-1536 binary system is shown.}
\label{fig:map}
\end{figure}

In the case of the 253-1536 system, the distance to the $\theta^1$ Ori C massive star, i.e. $\sim$ 1~pc in projection (see Fig.~\ref{fig:map}), is too large to make the gas in the disks being significantly ionized by the $\theta^1$ Ori C UV-radiation. This is indicated by the lack, in these ONC peripheral regions, of photoevaporating proplyds with bright ionization cusps pointing toward the direction of $\theta^1$ Ori C in the HST images \citep{Ricci:2008}. At the same time, a bright rim of ionized gas is visible in both the HST and EVLA maps on the east side of the binary system (Fig.~\ref{fig:map2}). The star which is causing this feature is probably NU~Orionis, an early B-type star located toward the direction of the bright rim, which is also responsable for the excitation of the M43 HII region \citep{Smith:2005}. 

To evaluate the relative impact of dust and ionized gas emission at a given frequency, the slope of the SED can be used: dust thermal emission presents a steep spectrum, with values of the spectral index $\alpha_{\rm{mm}}$ ($F_{\nu}\propto \nu^{\alpha}$) between about 2 and 3, whereas the ionized gas emission is characterized by a much more shallow dependence on frequency, e.g. $\alpha \simless 1$ for a symmetric, ionized, optically thick wind \citep{Panagia:1975}, and $\alpha \simless 0$ for optically thin emission \citep{Mezger:1967}. For the 253-1536 system, \citet{Ricci:2011} measured a spectral index of $2.6\pm0.4$ between 0.88 and 2.9~mm. Since the observations at 2.9~mm did not spatially resolve the binary system, this value is descriptive of the whole system rather than of the two disks separately. By adding up the 6.9~mm-flux densities for the two disks, the spectral index of the whole system is $2.0\pm0.5$ between 2.9 and 6.9~mm and $2.3\pm0.2$ between 0.88 and 6.9~mm. Both the measured absolute values of the spectral indeces, and the fact that no statistically-significant changes are seen in the slope of the spectrum between 0.88 and 6.9~mm suggest that the dust dominates the total sub-mm/mm emission. This is consistent with what found also for 216-0939, another HST-identified disk in the northern outskirts of the ONC \citep{Ricci:2011}. 
In the following we will therefore assume that the measured sub-mm/mm fluxes are only due to thermal emission from dust particles in the disks.

\begin{table*}
\centering \caption{Disk fluxes and spectral indeces.} \vskip 0.1cm
\begin{tabular}{lccccc}

\hline \hline
Source & $F_{\rm{6.9mm}}$ & rms$_{\rm{6.9mm}}$ & $F_{\rm{0.88mm}}$ & rms$_{\rm{0.88mm}}$ & $\alpha_{\rm{0.88-6.9mm}}$  \vspace*{1mm} \\
       &     (mJy)        &      (mJy)         &       (mJy)       &      (mJy)     &      \\
\hline

253-1536a & 1.10 & 0.035  & 135.0   & 1.0  & 2.34$\pm$0.16  \\
253-1536b & 0.34 & 0.035  & 38.4    & 1.0  & 2.30$\pm$0.19  \\
\hline
\end{tabular}
\begin{flushleft}
Column 6: Spectral index between 0.88 and 6.9~mm; reported errors take into account a 10~\%-uncertainty on the absolute flux scale at both 0.88 and 6.9~mm.
\end{flushleft}
\label{tab:obs}
\end{table*}

\section{Discussion}

\subsection{Comparison of grain growth in the two disks}
\label{sec:grain_growth}

\noindent 
Since the dust emission of disks at these long wavelengths is mostly optically thin, the spectral index $\alpha$ of the (sub-)mm SED is a proxy for the spectral index $\beta$ of the dust opacity coefficient ($\kappa_{\nu} \propto \nu^{\beta}$), which in turn carries the information on grain growth in the disk. For example, in the Rayleigh-Jeans regime for a completely optically thin emission, $\beta=\alpha-2$. 
However, since the hypotheses of completely optically thin emission in the Rayleigh-Jeans regime are not always realized even at these long wavelengths, modelling of the disk emission is required. 

We modelled the measured disk sub-mm SED of the 253-1536 binary system by using a modified version of the two-layer passively irradiated disk models \citep{Chiang:1997,Dullemond:2001}, as described in \citet{Ricci:2010a,Ricci:2010b}. 
The effective temperature of the two stellar companions were derived by converting the spectral types adopted in Section~\ref{sec:X-Shooter} with the temperature scale of \citet{Luhman:2003}. By using the \citet{Palla:1999} PMS evolutionary tracks and new multi-band optical photometry and spectroscopy, \citet{DaRio:2010} recently derived a new estimate for the ONC mean age of about 2~Myr. With these same PMS evolutionary tracks and age, the values for the stellar masses and luminosities are about 2~$M_{\odot}$, 12~$L_{\odot}$ for 253-1536a and about 0.3~$M_{\odot}$, 0.2~$L_{\odot}$ for 253-1536b, respectively.

The angular resolutions of the EVLA and SMA observations do not allow us to properly constrain the radial profile of the dust surface density in the two disks. For this reason, we consider in this analysis disks with truncated power-law surface densities ($\Sigma_{\rm{dust}} \propto r^{-p}$ for $r < R_{\rm{out}}$ and $\Sigma_{\rm{dust}}=0$ for $r > R_{\rm{out}}$) with possible $p$-values between 0 and 1.5, as obtained through high-angular resolution sub-mm imaging of disks in nearby SFRs \citep{Andrews:2007}. As for the disk outer radius $R_{\rm{out}}$, the HST and SMA observations constrained a value of $R_{\rm{out}} \approx 280$~AU for the 253-1536a disk and derived an upper limit of 60~AU for the unresolved 253-1536b disk. Since the $R_{\rm{out}}$ of the 253-1536b disk is not determined by the observations, in the following analysis we will consider two different possibilities, namely $R_{\rm{out}} = $40 and 60~AU. Smaller disks with $R_{\rm{out}} \simless 30~$AU always fail to reproduce the measured (sub-)mm fluxes of 253-1536b. This is due to the fact that the emission of such small and dense disks becomes optically thick and, as a consequence, underestimates the relatively large (sub-)mm fluxes of 253-1536b \citep[][]{Testi:2001}. Note that these possible values for the outer radius of the 253-1536b disk are all significantly lower than the estimated radius of the Roche lobe, i.e. $\simgreat 100$~AU~\citep[from ][using our estimates for the stellar masses, and the projected physical separation as a lower limit for the binary semi-major axis]{Paczynski:1971}. This means that the material in the disk lies well inside the stable zone in the Roche lobe.

Another important parameter is the disk inclination, defined as the angle between the disk axis and the line-of-sight. By taking the ratio of the two projected disk axes in the HST images, we estimated an inclination $i \sim 55^{\circ}$ for the 253-1536a disk. This procedure cannot be applied to the smaller 253-1536b disk, which has not been detected by HST. The X-Shooter spectra show that the 253-1536b star is significantly less extincted than its stellar companion. This indicates that the 253-1536b disk is likely less inclined than the companion. For this disk we considered a range of possible inclinations between 0$^{\circ}$ and 50$^{\circ}$.

With the parameters outlined above we determined the spectral index $\beta$ of the dust opacity coefficient by fitting the long-wave SED with the two-layer disk models~\citep[see][]{Ricci:2010a}.
Table~\ref{tab:models} shows the constrained $\beta$-values for the possible different $R_{\rm{out}}$ and $i$ for the 253-1536b disk, and in the $p=1$ case. The difference between the estimated $\beta$-values for the 253-1536b disk is given by the different contribution of the inner optically thick regions to the total emission. The decrease of $R_{\rm{out}}$, or the increase of $i$, which increases the line-of-sight optical depth of the disk, makes this contribution more important. As a consequence, $\beta$ becomes larger, i.e. the spectrum of the dust emissivity steepens, to compensate for the opposite effect given by optically thick emission. To better quantify the value of $\beta$ for the 253-1536b disk, high-angular resolution imaging is needed to directly constrain its outer radius, and therefore the impact of the optically thick inner regions to its total emission.  
The important point to be noticed here is that for all the possible values of the disk parameters (this result is unchanged for possible other $p$-values between 0 and 1.5), the $\beta$-value constrained for 253-1536b is larger than for 253-1536a.


\begin{table*}
\centering \caption{Disk parameters in the $p=1$ case.} \vskip 0.1cm
\begin{tabular}{lccccc}

\hline \hline
Source & $R_{\rm{out}}$ & $i$ & $\beta$ & $\Sigma_{\rm{gas,outer}}$ & $T_{\rm{dust,outer}}$   \vspace*{1mm} \\
       &     (AU)       & $(^{\circ})$ & &       (g/cm$^2$)         &      (K)       \\
\hline

253-1536a & 280 & 55 & 0.5$\pm$0.2 & $1.5 - 4$ & $18 - 29$   \vspace*{2mm}  \\ 
253-1536b & 40  & 0  & 1.0$\pm$0.2 & $30 - 60$ & $18 - 25$           \\
		  & 40  & 30 & 1.4$\pm$0.2 & $50 - 100$ & $18 - 25$        \\
		  & 40  & 50 & ...         & ... & ...        \\
		  & 60  & 0  & 0.7$\pm$0.2 & $10 - 20$ & $15 - 20$           \\
		  & 60  & 30 & 0.8$\pm$0.2 & $11 - 22$ & $15 - 20$        \\
		  & 60  & 50 & 0.9$\pm$0.2 & $12 - 24$ & $15 - 20$        \\

\hline
\end{tabular}
\begin{flushleft}
Columns 5,6: ranges of gas surface densities and dust temperatures constrained in the disk outer regions (Section~\ref{sec:results}). These values have been obtained by adopting the dust model described in \citet{Ricci:2010a}. However, note that for our analysis, only the ratio of the surface densities and dust temperatures in the two disks is relevant. 
A model with $R_{\rm{out}}=40~$AU and $i=50^{\circ}$ cannot reproduce the observed fluxes for 253-1536b.
\end{flushleft}

\label{tab:models}
\end{table*}

The $\beta$-index carries information on the size $a_{\rm{max}}$ of the largest grains in the dust population of the outer disk \citep[e.g.][]{Natta:2007}. $\beta$-values lower than about $1-1.5$ can only be explained with the presence of grains as large as at least $0.1-1~$mm. 
Converting an estimate for $\beta$ into one for $a_{\rm{max}}$ is particularly difficult because of our ignorance on the phsyical/chemical properties of the probed dust.
However, for all the dust models considered in the literature, a general anticorrelation between $\beta$ and $a_{\rm{max}}$ is generally obtained for $\beta \simless 1.5$.
Under the assumption that the chemistry/shape of the dust grains in the two disks is the same and
considering that the dust evolution models in disks with the constrained physical properties predict a slope $q$ which is nearly identical \citep{Birnstiel:2011}, this means that the observational data indicate that the 253-1536a disk, with a lower $\beta$, contains larger grains in its outer regions than 253-1536b.  
By considering the same dust model adopted in \citet{Ricci:2010a,Ricci:2010b,Ricci:2011}, and varying the assumed $q$-value between 2 and 3 \citep[see][]{Ricci:2010a}, the inferred $a_{\rm{max}}$ in the 253-1536a disk is larger than in the companion disk by factors of about $2-10$, depending on the values for $R_{\rm{out}}$ and $i$ for the 253-1536b disk.




\subsection{Testing the models of dust evolution}
\label{sec:results}

\noindent Recent physical models of dust evolution which include coagulation and fragmentation of dust grains predict a local relation 

\begin{equation}
a_{\rm{max}} = \frac{\Sigma_{\rm{gas}}}{\pi \alpha_{\rm{visc}}~c_s^2}\frac{u_{\rm{f}}^2}{\rho_s},
\label{eq:amax_til}
\end{equation}
between the maximum grain size allowed by fragmentation, the gas surface density $\Sigma_{\rm{gas}}$, the viscosity $\alpha_{\rm{visc}}$-parameter \citep{Shakura:1973}, the thermal speed $c_s$, the critical velocity $u_{\rm{f}}$ above which two grains fragment after colliding, and the grain density $\rho_s$ \citep[see][]{Birnstiel:2010a,Birnstiel:2010b}.

By assuming that the chemical composition and shape of dust grains in the two disks of the binary system is the same, the highly uncertain term $u_{\rm{f}}^2/ \rho_s$ in Eq.~\ref{eq:amax_til} cancels out when taking the ratio of $a_{\rm{max}}$ in the two disks:

\begin{equation}
\frac{a_{\rm{max,1}}}{a_{\rm{max,2}}} \approx \frac{\Sigma_{\rm{gas,1}}}{\Sigma_{\rm{gas,2}}}\frac{c_{s,\rm{2}}^2}{c_{s,\rm{1}}^2}\frac{\alpha_{\rm{visc,2}}}{\alpha_{\rm{visc,1}}} \approx \frac{\Sigma_{\rm{gas,1}}}{\Sigma_{\rm{gas,2}}}\frac{T_{\rm{dust,2}}}{T_{\rm{dust,1}}}\frac{\alpha_{\rm{visc,2}}}{\alpha_{\rm{visc,1}}},
\label{eq:amax_ratio}
\end{equation}
where $T_{\rm{dust}}$ is the dust temperature in the disk midplane. This means that observational constraints on the ratio of $a_{\rm{max}}$ in different disks allow to test models of dust evolution in disks without basing the whole analysis on parameters whose values are very uncertain.

Equations~\ref{eq:amax_til} and \ref{eq:amax_ratio} are valid locally in the disk. However, as noted above, our observations do not have enough angular resolution to properly resolve the disk structure and directly constrain the radial dependence of quantities like e.g. the dust surface density. The estimates obtained above on the dust properties in the two disks are referred to the disk regions which dominate the emission at long wavelengths, i.e. the outer regions.  
We can therefore attempt a comparison between the predictions of the dust evolution models and our observational results in the disk ``outer regions'', which have to be properly defined.

Since the two disks have very different outer radii, the spatial regions probed by their (sub-)mm SED are different. According to the two-layer disk models, more than 50\% of the total sub-mm emission from 253-1536a comes from regions with stellocentric radii $r > 100~$AU, whereas for 253-1536b the same fraction of emission comes from $r > 20-30~$AU. 
As shown in Table~\ref{tab:models} for the $p=1$ case, in these regions the 253-1536b disk is denser than 253-1536a by a factor of about $10-30$, and colder by a factor of about 1.2-1.4. By including these values in Eq.~\ref{eq:amax_ratio}, and assuming the same $\alpha_{\rm{visc}}$-value in the two disks, it is evident that the models of dust evolution would predict significantly larger grains in the 253-1536b disk, which is contrary to what was derived in the last section. This result does not change even when considering other possible $p$-values for the power-law index of the disk surface density between 0 and 1.5.

To reconcile the prediction of the models of dust evolution with the ratios of $a_{\rm{max}}$ reported in Section~\ref{sec:grain_growth}, the viscosity $\alpha_{\rm{visc}}$-parameter in the 253-1536b disk has to be larger than in the companion disk by more than a factor of 10. Physically, this is because to explain the smaller grains observed in the outer regions of the 253-1536b disk, the turbulence velocity, which is roughly proportional to $\sqrt{\alpha_{\rm{visc}}}$, has to be very high to increase the relative velocities between solid particles thus making fragmentation more efficient.

Although different $\alpha_{\rm{visc}}$-values could in principle be present in the two disks, magneto-rotational simulations of protoplanetary disks predict that larger values of $\alpha_{\rm{visc}}$ are typically obtained in environments with lower densities \citep{Gammie:1996}, which is the opposite of what requested by the dust evolution models to explain the observational results for the 253-1536 binary system.      
The fact that larger grains are seen in environments with lower densities probably suggests that radial motion of particles in the disks, a phenomenon which is not included when deriving this prediction, plays a fundamental role in the redistribution of solid particles in protoplanetary disks. This is also what \citet{Birnstiel:2010b} suggested to reconcile the predictions of these same models with the measured sub-mm fluxes of isolated disks in Taurus and Ophiuchus SFRs.  

High angular resolution and high sensitivity observations with ALMA and EVLA will allow us to test the dust evolution models locally in the disk, and possibly to probe viscosity with high spectral and spatial resolution observations of gas.

%
%

\acknowledgments

Based on observations collected at the European Southern Observatory, Chile. Program 284.C-5028.
We thank the ESO support astronomers in Chile for the observations conducted in service mode. 
L.R. acknowledges the PhD fellowship of the International Max-Planck-Research School.

\end{document}